\documentclass[final,5p,times,twocolumn]{elsarticle}
\usepackage{geometry}
\usepackage{csquotes}
\usepackage{graphicx}
\usepackage{xcolor}
\usepackage[hyphens]{url}
\usepackage{hyperref}
\usepackage{longtable}
\usepackage{amssymb}
\usepackage{multirow}
\usepackage{booktabs,siunitx} 
\usepackage{comment}
\usepackage{balance}
\usepackage{color}
\usepackage{ifthen}
\usepackage{enumerate}
\newboolean{showcomments}
\setboolean{showcomments}{true} 
\ifthenelse{\boolean{showcomments}}
    {
        \newcommand{\nauman}[1]{\textcolor{orange}{{\it [Nauman says: #1]}}}
        \newcommand{\usman}[1]{\textcolor{blue}{{\it [Usman says: #1]}}}
        \newcommand{\todo}[1]{\textcolor{red}{{\it [TODO: #1]}}}
        \newcommand{\claes}[1]{\textcolor{cyan}{{\it [Claes says: #1]}}}
    }
    {
        \newcommand{\nauman}[1]{}
        \newcommand{\usman}[1]{}
        \newcommand{\claes}[1]{}
        \newcommand{\todo}[1]{}
    }

\begin{document}

\journal{Information and Software Technology}
\begin{frontmatter}
\title{A Quality Assessment Instrument for\\ Systematic Literature Reviews in Software Engineering}
		
\author{ Muhammad Usman, Nauman bin Ali, Claes Wohlin}
\address{Blekinge Institute of Technology, Karlskrona, Sweden \\
	\{muhammad.usman, nauman.ali, claes.wohlin\}@bth.se
}

\begin{abstract}
\textbf{Context:} Systematic literature reviews (SLRs) have become standard practise as part of software engineering research, although their quality varies. To build on the reviews, both for future research and industry practice, they need to be of high quality. \textbf{Objective:} To assess the quality of SLRs in software engineering, we put forward an appraisal instrument for SLRs. The instrument is intended for use by appraisers of reviews, but authors may also use it as a checklist when designing and documenting their reviews. \textbf{Method:} A well-established appraisal instrument from research in healthcare was used as a starting point to develop a quality assessment instrument. It is adapted to software engineering using guidelines, checklists, and experiences from software engineering. As a validation step, the first version was reviewed by four external experts on SLRs in software engineering and updated based on their feedback. \textbf{Results:} The outcome of the research is an appraisal instrument for quality assessment of SLRs in software engineering. The instrument intends to support the appraiser in assessing the quality of an SLR. The instrument includes 16 items with different options to capture the quality. The item is assessed on a two or three-grade scale, depending on the item. The instrument also supports consolidating the items into groups, which are then used to assess the overall quality of a systematic literature review. \textbf{Conclusion:} It is concluded that the presented instrument may be helpful support for an appraiser in assessing the quality of SLRs in software engineering.

\end{abstract}


\end{frontmatter}


\section{Introduction}
\label{sec:intro}
To establish evidence-based practices in software engineering (SE), Kitchenham~\cite{kitchenham2004procedures} proposed the use of systematic literature reviews (SLRs) in software engineering to identify, appraise and synthesise evidence reported in the scientific literature. Today the method is well-accepted in SE. This is illustrated by the growing number of published SLRs. The number of SLRs has increased rapidly since the introduction of the guidelines~\cite{AliU18}. Furthermore, several individual researchers and research groups (beyond the proposers of the guidelines) are actively conducting and publishing SLRs. Moreover, SLRs are published in almost all top-tier SE journals and conferences and cover various topics that encompass all knowledge areas of SE. In summary, SLRs have become a standard practice in SE. A standard practice comes with requirements on, for example, the reliability of SLRs.

Thus, the reliability of conducting SLRs as a method needs to be maintained, ensuring trust in the results. Several researchers have proposed guidelines and checklists to design, conduct and report SLRs. However, relatively little work has been done on the critical appraisal of the SLRs. Several researchers have replicated SLRs or applied appraisal instruments to assess the quality of SLRs. Replications of SLRs are a resource-intensive undertaking, i.e., it is not scalable. Some replications exist, for example, some researchers have relied on intentional replication~\cite{MacDonellSKM10}, and other researchers have taken the opportunity to compare SLRs when two SLRs on the same topic were published~\cite{WohlinRNEMA13}. In both cases, the objective was to evaluate the reliability of SLRs. An alternative to expensive replication-based evaluation is the use of critical appraisal instruments. Such instruments help assess the quality of an SLR by checking its conformance with best practices and established guidelines. Several SE researchers have used an interpretation of the criteria used by the Centre for Reviews and Dissemination (CRD) at the University of York to include an SLR in their Database of Abstracts of Reviews of Effects (DARE)~\cite{DARE}. However, these questions are insufficient to reveal significant limitations in SLRs as noted in our previous work~\cite{AliU18}.

The adherence to design and reporting guidelines helps to improve the reliability and transparency of SLRs. However, these guidelines do not provide the user the means to make a critical judgement of the risk of bias that certain decisions and actions in the design and execution of SLRs may introduce. We exemplify this distinction between the role of reporting checklists and critical appraisal tools with the following example~\cite{ALI201948}: a reporting checklist will only ask if a specific aspect is described, e.g., \textit{``Are the review’s inclusion and exclusion criteria described?''}, whereas a critical appraisal tool would also raise the question of appropriateness. In the case of selection criteria, an appraisal tool poses the question \textit{``Are the review’s inclusion and exclusion criteria appropriate?''}. Thus, an appraisal tool also considers the quality, i.e. not only that a certain item is present.

In evidence-based medicine, several critical appraisal instruments have been developed and evaluated beyond the DARE criteria. In our previous work, we identified AMSTAR~2 (A MeaSurement Tool to Assess systematic Reviews)~\cite{Sheaj4008} as a candidate tool for adoption in SE. We concluded that the tool would need adaptation for SE. In this paper, we report the process and outcome of adapting AMSTAR~2 for the quality appraisal of SLRs in SE. We call the adapted instrument \textbf{Q}uality \textbf{A}ssessment \textbf{I}nstrument for \textbf{S}oftware \textbf{E}ngineering systematic literature \textbf{R}eviews (abbreviated as QAISER). 

Our approach when developing QAISER has several salient features focusing on increasing the reliability of the research outcome. We based our work on a well-accepted and validated instrument as a foundation (i.e., AMSTAR~2)~\cite{Sheaj4008}. To ensure an appropriate adaptation to SE, we collected and relied on a comprehensive set of documents with guidelines and best practices for conducting SLRs in SE. We followed a systematic and well-documented process to develop QAISER with several internal validation steps involving multiple researchers. Furthermore, we invited some leading experts in evidence-based SE research to conduct an external validation of QAISER. In each step of the process, QAISER was updated based on the feedback received and internal discussions. The outcome of the process, i.e., QAISER, is the main contribution of the paper.

The remainder of the paper is organised as follows. Section~\ref{sec:related} presents an overview of the main critical appraisal instruments used in both SE and evidence-based medicine. Section~\ref{sec:method} describes in detail the method undertaken for developing QAISER. In Section~\ref{sec:Conduct}, it is described how QAISER evolved into the latest version, which is the main outcome of the research presented. Section~\ref{sec:results} presents the QAISER appraisal instrument for SLRs in SE in detail. In Section~\ref{sec:reliability} we reflect on the reliability of QAISER. Section~\ref{sec:application} describes the guidance document and shares our reflections about using QAISER on three example SLRs. The threats to validity are presented in Section~\ref{sec:ValidityThreats}. Section~\ref{sec:discussion} discusses the implication of the results. Section~\ref{sec:conclusion} concludes the paper and presents a roadmap for further evaluation and our ambition to support broader adoption of QAISER. Finally, the QAISER instrument is provided in Appendix A (attached as supplemental material), and a guidance document supporting the instrument can be found in Appendix B (also attached as supplemental material).

\section{Related work}
\label{sec:related}

A prerequisite for a quality appraisal is that we pose the right questions. In the first version of the guidelines for systematic literature reviews in SE, Kitchenham~\cite{kitchenham2004procedures} identified two sets of questions from Greenhalgh~\cite{Greenhalgh1997} and Khan et al.~\cite{Khan2001} to review any existing SLRs on a topic of interest. In the 2007 update~\cite{keele2007guidelines}, Kitchenham and Charters added the CRD Database of Abstracts of Reviews of Effects (DARE) set of four questions to the list~\cite{DARE}. Kitchenham and Charters~\cite{keele2007guidelines} also applied the criteria to SLRs published between 2004 and 2007. 

The proposal from Greenhalgh is very general; Khan et al.'s proposal is the most comprehensive, while the DARE criteria are brief and ``simple''~\cite{keele2007guidelines}. Among these three sets of questions proposed in the guidelines, only the DARE criteria have been widely used in the SE literature.

Kitchenham et al.~\cite{kitchenham_systematic_2010} provided guidance to answer four of the five questions in the DARE criteria. Cruzes and Dyb{\aa}~\cite{CRUZES2011440} observed that one of the critical questions regarding synthesis had not been included in the SE guidelines for conducting SLRs and has not been used when evaluating the quality of SLRs in SE. It should be noted that the number of questions in DARE has varied over the years; it has included either four or five questions depending on the version of DARE. 

Some others have developed their own interpretation of the DARE questions~\cite{NurdianiBF16,AliPW14}. One shared limitation of these is the lack of traceability between the proposals and the evidence/best practices used to motivate them. 
 
Other researchers have also been concerned with assessing quality in SLRs in SE, Dyb{\aa} and Dings{\o}yr~\cite{dyba_strength_2008} reviewed several proposals from evidence-based medicine to assess the quality of SLRs. They concluded that the MOOSE statement~\cite{stroup2000meta} is a very relevant reporting checklist for SLRs in SE. The MOOSE checklist has six main reporting items including `background', `search strategy', `method', `results', `discussion' and `conclusions'. Each item further lists actions and details that should be provided in an SLR. 

In a previous study~\cite{ALI201948}, we reviewed the proposals for quality assessment for SLRs both from SE and other fields. We concluded that in the SE literature, there is an awareness of reporting checklists like MOOSE, QUOROM, and PRISMA. However, SE researchers have not yet leveraged the progress in critical appraisal tools for systematic reviews.

One essential aspect related to quality assessment is the validity threats presented by authors of SLRs. Ampatzoglou et al.~\cite{AMPATZOGLOU2019201} reviewed 100 secondary studies in SE and identified the commonly reported threats to validity and the corresponding mitigation actions. They also proposed a checklist that authors can use to design an SLR with explicit consideration for common validity threats and develop an informed plan for mitigating them. The authors state that readers can also use the checklist to assess the validity of the results of an SLR. The checklist has 22 questions grouped into three categories: study selection validity, data validity, and research validity. Furthermore, for each of the 22 questions, there are 1 to 9 sub-questions. 

The checklist by Ampatzoglou et al.~\cite{AMPATZOGLOU2019201} encapsulates the current state of research regarding mitigating validity threats. Also, the checklist is a useful design tool to support the design and execution of an SLR. However, we argue that it is not a tool that enables the evaluation of completed SLRs. In this study, we have used their work to develop QAISER.

Given the lack of an appraisal tool adapted for SE, we wanted to leverage experiences from other research fields. Through an analysis of the leading appraisal tools, including ROBIS, AMSTAR, and AMSTAR~2, we identified AMSTAR~2 (A MeaSurement Tool to Assess systematic Reviews)~\cite{Sheaj4008} as a candidate tool for adaptation to SE~\cite{ALI201948}. AMSTAR was developed based on a review of available rating instruments and consolidated them into 11 appraisal items. It has since been extensively used and validated. AMSTAR~2 is a revised version of the tool that takes into account the systematically collected community feedback. The major updates for AMSTAR~2 are: (1) the consideration of SLRs that may include non-randomized studies and (2) an increased focus on the risk of bias evaluation.

\begin{figure*}[ht]
    \includegraphics[width=0.95\textwidth]{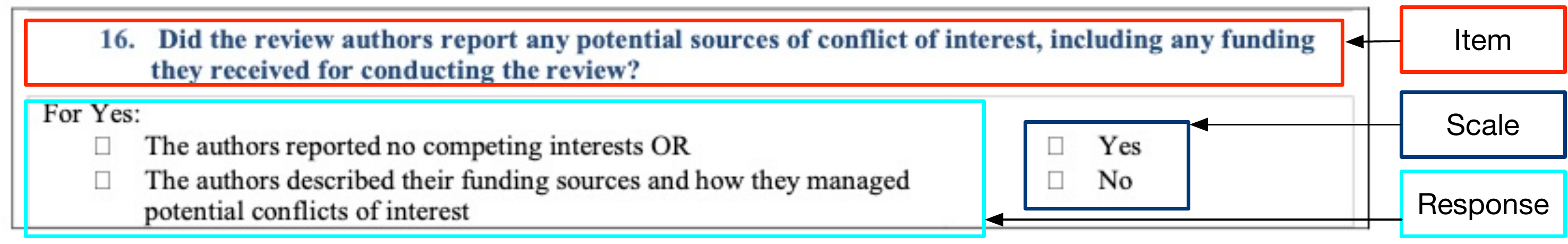}
    \centering
    \caption{Items, responses and scale in AMSTAR~2.} 
    \label{fig:AMSTAR_Key}
\end{figure*}

AMSTAR~2 provides more comprehensive coverage of important quality aspects of an SLR that are not included in the DARE criteria that are mostly used in SE~\cite{ALI201948}.
AMSTAR~2 consists of 16 appraisal items and their corresponding response options and scale. Figure~\ref{fig:AMSTAR_Key} annotates an example of an item, response, and scale from AMSTAR~2.

Based on an analysis of related work it was decided to use AMSTAR~2 as a basis for proposing a quality assessment instrument tailored for SE. 

\section{Method}
\label{sec:method}

\begin{figure}[ht]
	\includegraphics[width=0.4\textwidth]{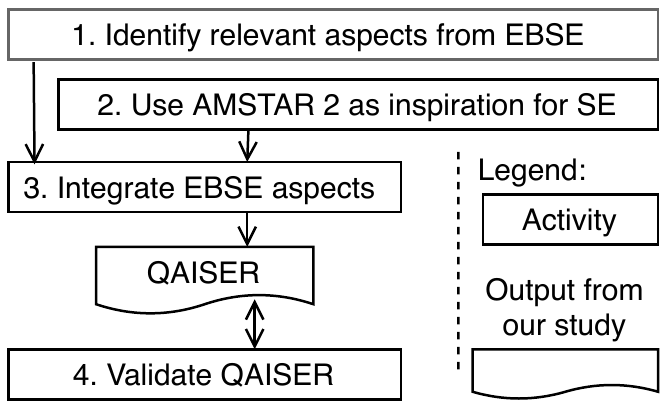}
	\centering
	\caption{Overview of the QAISER development process.} 
	\label{fig:QAISER_process}
\end{figure}

This section describes the four-step process we used to develop QAISER (see Figure~\ref{fig:QAISER_process} for an overview). In the first step, we identified aspects from the evidence-based software engineering (EBSE) literature relevant for inclusion in QAISER. In the second step, we adapted AMSTAR~2 for SE by customizing its items and responses. In the third step, we combined the outputs of the previous two steps by integrating the EBSE aspects into QAISER. Finally, in the fourth step, we validated QAISER by inviting external experts to evaluate its completeness, understandability, and relevance of its items and responses for SE.

\begin{figure*}[ht]
	\includegraphics[width=1\textwidth]{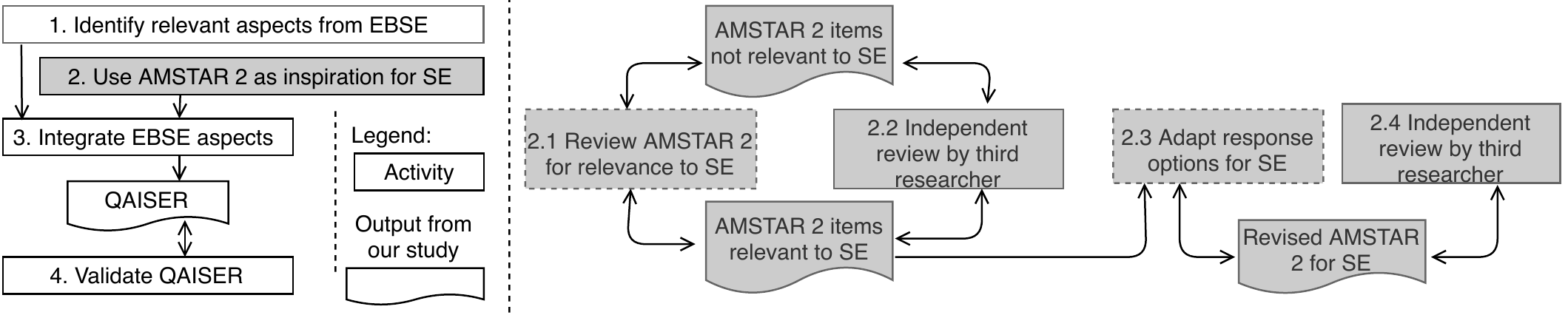}
	\centering
	\caption{Step 2: Using AMSTAR 2 as a source of inspiration for SE.} 
	\label{fig:Step2}
\end{figure*}

\begin{figure*}[ht]
	\includegraphics[width=1\textwidth]{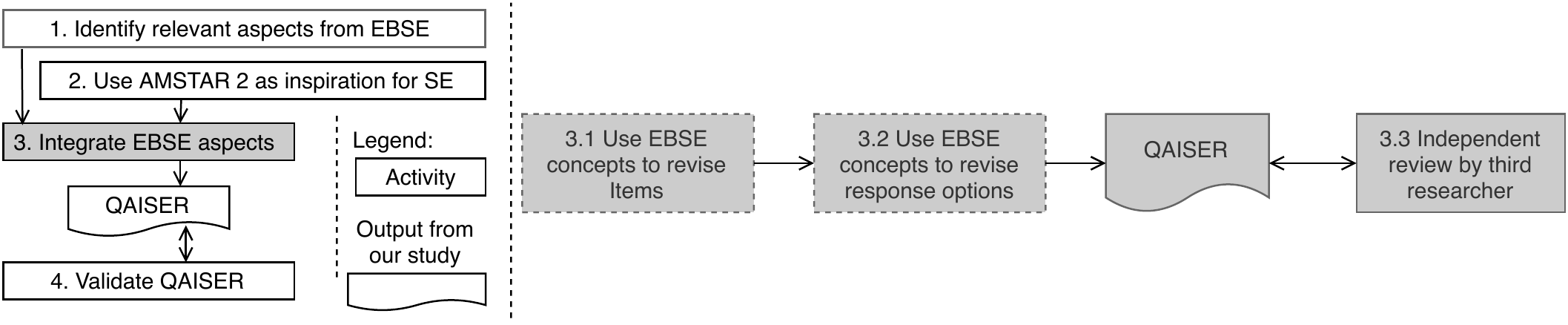}
	\centering
	\caption{Step 3: Integrating EBSE aspects.}
	\label{fig:Step3}
\end{figure*}

Each step is further elaborated below and the details of each step are also illustrated in Figures~\ref{fig:Step1} --~\ref{fig:Step4}. 

\subsection*{Step 1: Identifying relevant aspects from the EBSE literature}
\begin{figure*}[ht]
	\includegraphics[width=1\textwidth]{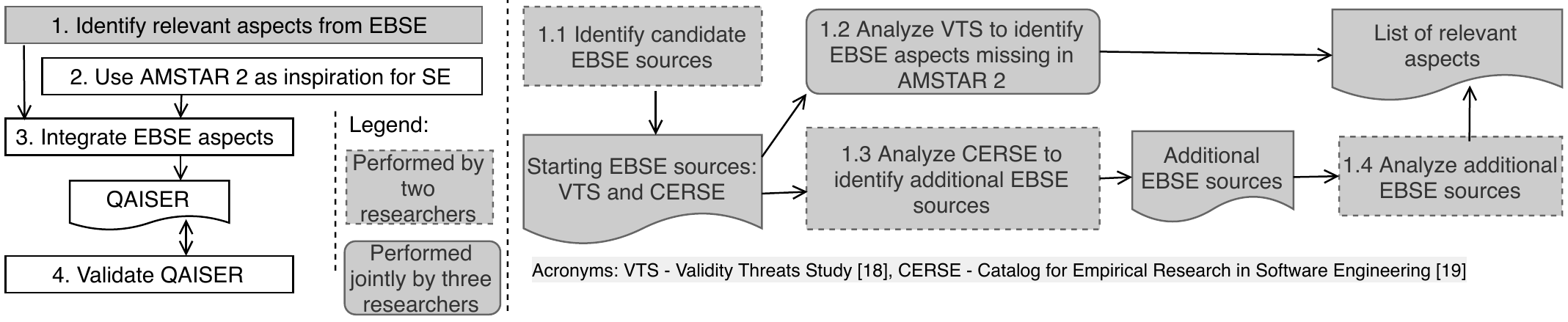}
	\centering
	\caption{Step 1: Identifying relevant aspects from EBSE literature.} 
	\label{fig:Step1}
\end{figure*}

In this step, we aimed to complement AMSTAR~2 with the relevant work from the EBSE literature. We followed a systematic approach to identify and analyze the relevant EBSE work (see Figure~\ref{fig:Step1}).

We started with analyzing a closely related and recent tertiary study on validity threats in SLRs in software engineering by Ampatzoglou et al.~\cite{AMPATZOGLOU2019201}. They have aggregated validity threats and corresponding mitigating actions in the form of a checklist as described above in Section~\ref{sec:related}. We analyzed their checklist to identify aspects that are covered or missing in AMSTAR~2~\cite{Sheaj4008}.

Molleri et al.~\cite{molleri2019cerse} recently proposed a Catalog for Empirical Research in Software Engineering (CERSE) based on a systematic mapping study of 341 methodological papers that were identified using a combination of manual and snowballing search strategies. CERSE includes available guidelines, assessment instruments, and knowledge organization systems for empirical research in software engineering. To identify additional relevant articles that are not covered by Ampatzoglou et al.~\cite{AMPATZOGLOU2019201} in their tertiary study, we selected 74 articles from CERSE that are related to systematic literature reviews (SLR) and mapping studies (SMS). We obtained the source file containing the basic information (title of the paper, publication venue etc.) for these 74 articles from the first author of CERSE~\cite{molleri2019cerse}. The first two authors independently reviewed these 74 articles to identify studies that propose or evaluate guidelines for conducting SLRs and SMSs in SE. Later, in a meeting, the first two authors developed a complete consensus on all 74 studies. The list of identified studies included, besides others, the latest version of the guidelines by Kitchenham et al.~\cite{kitchenham2015evidence}, the guidelines for mapping studies by Petersen et al.~\cite{PETERSEN20151} and the guidelines for snowballing by Wohlin~\cite{Wohlin:2014:GSS:2601248.2601268}. After including these three guidelines in our list of additional EBSE sources, we removed studies that were already covered in these guidelines~\cite{kitchenham2015evidence,PETERSEN20151,Wohlin:2014:GSS:2601248.2601268}.

\subsection*{Step 2: Using AMSTAR~2 as a source of inspiration for SE}
The first two authors jointly analyzed AMSTAR~2 to identify items that are relevant for SE. As a validation, the third author independently reviewed the list of relevant and non-relevant items identified by the first two authors. Next, the first two authors adapted the response options for SE, for example, by replacing the medicine-specific options with the appropriate SE options. The adapted response options were also reviewed independently by the third author. After discussions, we achieved complete consensus between all three authors on all changes in items and response options.

\subsection*{Step 3: Integrating EBSE aspects}
Using the outputs of the previous steps and, in particular, the relevant EBSE literature identified in Step 1, the first two authors developed the first draft of QAISER. They also prepared a guidance document to support QAISER users in applying the instrument. The third author independently reviewed the instrument and the guidance document to validate its contents, i.e., to check that any relevant aspect is not missed. The independent review helped improve the formulations and remove some inconsistencies in the instrument and the guidance document. However, it did not result in any significant change in the instrument.

\subsection*{Step 4: Validating QAISER}
\begin{figure*}[ht]
	\includegraphics[width=1\textwidth]{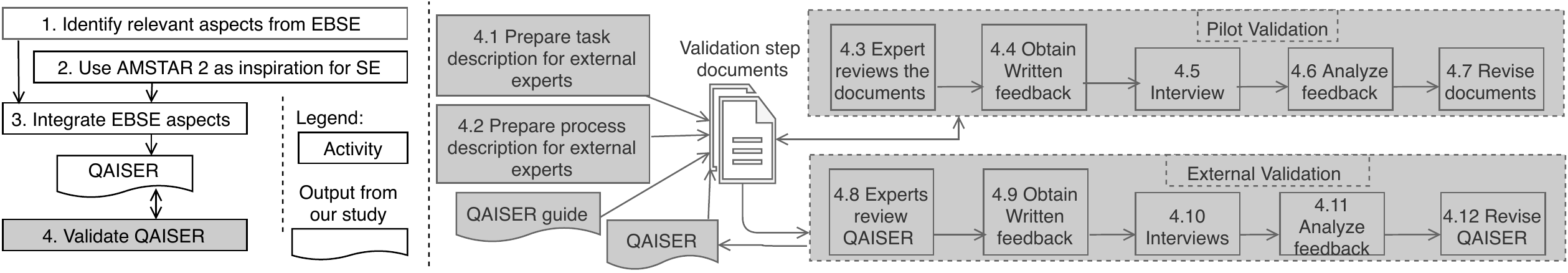}
	\centering
	\caption{Step 4: Validating QAISER.} 
	\label{fig:Step4}
\end{figure*}

In this step, QAISER was reviewed by four leading experts in EBSE to validate the appropriateness of its items and reflect on its completeness (i.e., to identify if some aspects are missing) and understandability. In addition to QAISER and the guidance document, we prepared the following two documents to conduct the validation step (see Figure~\ref{fig:Step4} for details about the validation step) systematically:

\begin{itemize}
    \item A task description document: It described the steps that the external experts were asked to perform while reviewing QAISER. The task description document provided space where experts could enter their feedback on each QAISER item.
    \item A process description document: It briefly described the process we used to create QAISER.
\end{itemize}

Before the external validation, we performed a pilot validation with a senior colleague at our department who has experience of participating in multiple SLRs. The colleague reviewed all of the four documents mentioned above (i.e., task description, process description, QAISER, and the guidance document) and provided written feedback. We also conducted a follow-up interview (one hour, face-to-face) to discuss the feedback in detail and to ensure a shared understanding. We revised the task description and also the instrument based on the feedback collected during the pilot step. Most of the changes resulted in revised formulations. We shared the revised documents with our colleague and achieved consensus on the clarity of the task description and completeness and appropriateness of QAISER.

Next, we used the same approach with the external experts as we followed during the pilot. After obtaining the written feedback and performing the interviews (approximately one hour each and on distance) with all four external experts, we analyzed the comments to identify the changes that should be made in QAISER. Also, a revised version of QAISER (the one presented in Appendix A in the supplemental material) and a summary of their feedback and our actions were sent to the external experts.

\section{Details of conducting the study}
\label{sec:Conduct}

In this section, we present details of how we applied the process described in Section~\ref{sec:method} and our justifications for the proposed changes to AMSTAR~2 while adapting it for SE. 

\subsection{Development of QAISER~V0}
\label{subsec:QAISERV0}

In Step 1 of the process described in Section~\ref{sec:method}, we identified and selected four sources (see~\cite{AMPATZOGLOU2019201,kitchenham2015evidence,PETERSEN20151,Wohlin:2014:GSS:2601248.2601268}), in addition to DARE~\cite{DARE}, from the EBSE literature to identify the relevant aspects for QAISER. Later, based on the suggestions of the external experts, we also included two more sources for identifying the relevant aspects for QAISER. The two additional sources related to a framework for an automated-search strategy to improve the reliability of searches in SLRs~\cite{ALI201948}, and a tertiary study describing lessons learned about reporting SLRs~\cite{BudgenBDW18}.

We now present the adaptation of AMSTAR~2 for SE based on the procedure detailed in Steps 2 and 3 of our method (see Section~\ref{sec:method}). Overall, at this stage in the process, we had two major changes in AMSTAR~2. The first change relates to the removal of existing items in AMSTAR~2. The removal includes excluding one item and replacing two items with a general item that is more appropriate for SE. The second change concerns the addition of an item.

In terms of removed items, three AMSTAR~2 items were not included in QAISER as these were not deemed relevant to SLRs in SE. The details for Items 1, 11 and 12 are described later in the section. However, in summary, AMSTAR~2 Item 1 is about using PICO (Population, Intervention, Comparator group, Outcome) components in research questions and selection criteria. While Items 11 and 12 are about meta-analysis, which is not commonly conducted in SE SLRs. We replaced these two items with a more general item about synthesis (see QAISER Item 11 in Appendix A). The new item checks if the included studies are synthesized or not. Synthesis of the included studies is one of the essential steps in an SLR~\cite{AMPATZOGLOU2019201,BudgenBDW18,DARE,kitchenham2015evidence}.

The addition of one item is due to the following. Item 5 in AMSTAR~2 checks if the study selection is performed independently by at least two authors of the review. Item 6 checks the same aspect about the data extraction process. However, no item in AMSTAR~2 checks if the quality assessment is performed independently by at least two persons. We introduced an additional item to cover this aspect, i.e., to see if the quality assessment is performed independently by at least two authors of the review (see QAISER Item 10 in Appendix A).

We now describe in detail why and what changes were made to each item in AMSTAR~2. \newline 

\noindent\textbf{Item 1. \textit{\enquote{Did the research questions and inclusion criteria for the review include the components of PICO?}}} \newline 
In SE, researchers are not widely using PICO when developing research questions and inclusion criteria. It is also not part of the revised guidelines~\cite{kitchenham2015evidence}. \newline \underline{\textit{Changes:}} This item is not relevant for SE and was excluded from QAISER. \newline 

\noindent\textbf{Item 2. \textit{\enquote{Did the report of the review contain an explicit statement that the review methods were established prior to the conduct of the review and did the report justify any significant deviations from the protocol?}}} \newline
We identified no need to make any change at the item level. However, the following issues in the response options were noted:
\begin{enumerate}[a)]
    \item The response options for 'Partial Yes' lack several aspects that are part of the protocol template included in the revised guidelines~\cite{kitchenham2015evidence}. The missing aspects include description of the need for the review, data extraction process, synthesis process, threats to validity of the review, deviations from the protocol and the corresponding justifications for such deviations, and details of conducting the review. 
    \item Under 'Partial Yes', the authors are only required to state that they had a written protocol. The authors should also make the protocol publicly accessible and describe where and how can it be accessed~\cite{kitchenham2015evidence}.
    \item One of the response options uses the term \enquote{risk of bias assessment}. In SE, the more commonly used term is quality assessment.
\end{enumerate}
\underline{\textit{Changes:}} Based on the analysis, the response options were modified as follows as an adaptation of them for SE:
\begin{enumerate}[a)]
\item The missing response options under 'Partial Yes' were added. 
\item In the revised item, the authors are also required to make the protocol accessible and state how and where it can be accessed. 
\item The risk of bias related response option was rephrased as quality assessment. \newline
\end{enumerate}

\noindent\textbf{Item 3. \textit{\enquote{Did the review authors explain their selection of the study designs for inclusion in the review?}}} \newline
Most reviews in SE include different types of empirical studies. Thus, it is not relevant to ask for a justification for including all types of study designs. Furthermore, the study design is only one of the criteria for including or excluding studies from an SLR. Therefore, the item should address the larger aspect of the appropriateness of the inclusion and exclusion criteria. Reporting of the inclusion and exclusion criteria is also part of the criteria used by the Centre for Reviews and Dissemination at the University of York to include an SLR in their Database of Abstracts of Reviews of Effects (DARE)~\cite{DARE}. Also, reporting of the inclusion and exclusion criteria and the relevant justifications is part of the guidelines~\cite{kitchenham2015evidence} and other EBSE literature as well~\cite{AMPATZOGLOU2019201}. \newline \underline{\textit{Changes:}} The item is revised as follows for SE: \textit{Are the review's inclusion and exclusion criteria appropriate?}
\newline To get a 'Yes' score for the revised item, the review should have reported the inclusion and exclusion criteria and provided justifications for any restrictions used in the criteria. \newline

\noindent\textbf{Item 4. \textit{\enquote{Did the review authors use a comprehensive literature search strategy?}}} \newline
We identified the following issues in the response options:
\begin{enumerate}[a)]
\item The response options treat database search as the main search method while snowballing is partially addressed as an additional search method. In the revised guidelines for performing SLRs in SE~\cite{kitchenham2015evidence}, database and snowballing searches are included as alternate search strategies. Both strategies have been used in SE SLRs and have their own guidelines and best practices (cf.\cite{Wohlin:2014:GSS:2601248.2601268,AliU18,kitchenham2015evidence}). In the current form, only the database search strategy could be assessed as comprehensive.
\item The response option related to the publication restrictions is more relevant to the inclusion and exclusion criteria.
\item Furthermore, two other response options are not used in SE SLR: The first one is about searching in the study registries, while the second one is about conducting the search within 24 months of completion of the review.
\end{enumerate}

\noindent \underline{\textit{Changes:}} We introduced the following three changes:
\begin{enumerate}[a)]
\item Two groups of response options were created: first when a database search is used as the main search method and the second when a snowballing search is used as the main search method (See QAISER Item 4 in Appendix A) for details about the two groups of response options).
\item The response option related to the publication restrictions is moved to Item 3 (see Appendix A).
\item The two response options (searching in registries and search within last 24 months) were not included in QAISER. \newline
\end{enumerate}

\noindent\textbf{Item 5. \textit{\enquote{Did the review authors perform study selection in duplicate?}}} \newline
We noted that:
\begin{enumerate}[a)]
\item The phrase \enquote{in duplicate} is not a commonly used term in SE and is therefore not self-explanatory. Furthermore, the item does not specify if the study selection is performed on the full text or on the titles and abstracts.
\item In the first response option, when all studies are reviewed independently by at least two authors of the review, the agreement level is not reported. Reporting of the agreement level would increase the transparency of the study selection process. 
\item In the second response option, it is permitted that only a sample of the studies are independently reviewed by at least two authors of the review. The reliability of the study selection process is compromised if only a small sample of studies is reviewed by more than one author of the review. In particular, the excluded studies pose a threat to validity if a single person excludes them.
\end{enumerate}
\noindent \underline{\textit{Changes:}} Three changes were introduced to address these observations:
\begin{enumerate}[a)]
    \item The item was rephrased to clarify the focus on the independent study selection and that the initial study selection is based on titles and abstracts. The revised formulation is: \textit{\enquote{Did the authors of the review \textbf{independently} perform study selection \textbf{based on titles and abstracts}?}} 
    \item At the end of the first response option, the following text is added to make it necessary to report the agreement level as well: \textit{\enquote{... and reported the agreement level}}. 
    \item At the end of the second response option, the following text is added to make it compulsory to have the excluded studies reviewed by at least two authors: \textit{\enquote{however, all excluded studies must be reviewed by at least two authors of the review}}.\newline
\end{enumerate}

\noindent\textbf{Item 6. \textit{\enquote{Did the review authors perform data extraction in duplicate?}}} \newline
\noindent As in the previous item, the phrase \enquote{in duplicate} is not self-explanatory. \newline \underline{\textit{Changes:}} The item was rephrased in QAISER as follows: \textit{\enquote{Did at least two authors of the review independently perform data extraction?}} \\

\noindent\textbf{Item 7. \textit{\enquote{Did the review authors provide a list of excluded studies and justify the exclusions?}}} \newline
The item is about those studies that were excluded after reading the full text. The item does not indicate that it is about those studies that were read in full text, and not about those that were excluded based on the screening of the titles and abstracts. \newline
\underline{\textit{Changes:}} The item was rephrased to indicate that it is about those studies that were read in full text. In the revised formulation, the following phrase is added at the end of the item text: \textit{\enquote{...for the papers read in full text?}}\\

\noindent\textbf{Item 8. \textit{\enquote{Did the review authors describe the included studies in adequate detail?}}} \newline
We did not note any issues in the item. However, the response options about intervention and outcomes may not be relevant to all SLRs in SE. In SE, not all SLRs would be about interventions and outcomes. The included studies in an SLR may not have investigated any interventions.  \newline
\underline{\textit{Changes:}} In the response options about interventions and outcomes, the phrase \textit{\enquote{when applicable}} is added to explain that the review needs to describe only the relevant information about included studies. \\

\noindent\textbf{Item 9. \textit{\enquote{Did the review authors use a satisfactory technique for assessing the risk of bias (RoB) in individual studies that were included in the review?}}} \newline
We noted the following:
\begin{enumerate}[a)]
\item In SE SLRs, the concept of quality assessment is used, instead of RoB, to refer to the quality assessment of the individual studies. A variety of quality assessment instruments have been developed and used to assess the quality of the different types of empirical studies in software engineering~\cite{kitchenham2015evidence}. The focus in SE is on using relevant quality assessment instruments.
\item The current response options are not relevant to SE. Furthermore, the focus of the item is suggested to be changed to the quality assessment instrument. Therefore, the response options should also be revised accordingly to check the completeness and relevance of the questions in the quality assessment instrument.
\end{enumerate}
\underline{\textit{Changes:}} We introduced the following changes:
\begin{enumerate}[a)]
\item We revised the item to emphasize whether or not the review authors have provided an explanation for their selection of the quality assessment instrument. The item is revised as follows: \textit{\enquote{Did the review authors explain their selection of quality assessment instrument?}}
\item With regards to the response options under the revised item, for 'Yes', the review authors should have selected an appropriate quality assessment instrument for different types of studies included in the review. Furthermore, the instrument needs to have questions about study goals, research questions, appropriateness of the study design, data collection, and analysis methods. The instrument should also have question(s) about the study findings and the supporting evidence, and the extent to which the findings answer the research questions. We refer to the instrument in Appendix A for the specific response options for this item in QAISER.\\
\end{enumerate}

\noindent\textbf{Item 10. \textit{\enquote{Did the review authors report on the sources of funding for the studies included in the review?}}} \newline
This item focuses only on the sources of funding for individual studies. Funding is one of the issues that could result in a conflict of interest. In some cases, the authors of the individual studies might have some other conflict of interest in favor of or against the topic or intervention they are investigating in their studies. \newline
\underline{\textit{Changes:}} The item is revised to include any other conflict of interest besides funding sources. Conflict of interest is inserted in the item text as follows: \textit{\enquote{Did the review authors report on the sources of funding \textbf{and any other conflict of interest} for the studies included in the review?}} \newline

\noindent\textbf{Item 11. \textit{\enquote{If meta-analysis was performed did the review authors use appropriate methods for statistical combination of results?}}} \newline
Meta-analysis studies are very rare in software engineering due to the lack of randomized controlled trials. Therefore, this item is not relevant to the majority of the SE SLRs. \newline
\underline{\textit{Changes:}} This item is removed from the adaptation of AMSTAR~2 for SE. We have instead included a more general item about synthesis (Item 11 in QAISER, see Appendix A).\newline

\noindent\textbf{Item 12. \textit{\enquote{If meta-analysis was performed, did the review authors assess the potential impact of RoB in individual studies on the results of the meta-analysis or other evidence synthesis?}}} \newline
As discussed with Item 11 above, meta-analysis is not common in SE SLRs. This item is removed from the adaptation of AMSTAR~2 for SE. However, it is important to note that considering the impact of the quality of individual studies while interpreting the results is still covered in the next item. \newline

\noindent\textbf{Item 13. \textit{\enquote{Did the review authors account for RoB in individual studies when interpreting/ discussing the results of the review?}}} \newline
We noted the following:
\begin{enumerate}[a)]
    \item Instead of RoB, the SE community uses the notion of quality assessment more commonly. 
    \item The first response option deals with the inclusion of high-quality randomized controlled trials (RCTs). Since in SE, RCTs are not common, the focus should be on high-quality studies.
    \item The second response option includes the requirement of discussing the impact of RoB on results. For SE, the focus has been on categorizing the analysis and interpretation of results based on study quality~\cite{kitchenham2015evidence}.
\end{enumerate}

\noindent \underline{\textit{Changes:}} The following changes were introduced:
\begin{enumerate}[a)]
    \item In the item description, the RoB is replaced with quality of individual studies.
    \item In the first response option, the phrase \textit{\enquote{high quality RCTs}} is replaced with \textit{\enquote{high quality studies}}.
    \item The second response option is revised to focus on the categorization of the analysis and interpretation of results based on study quality. \newline
\end{enumerate}

\noindent\textbf{Item 14. \textit{\enquote{Did the review authors provide a satisfactory explanation for, and discussion of, any heterogeneity\footnote{Heterogeneity occurs when the results are not consistent across studies. For example, different studies provide conflicting evidence for or against a software engineering intervention. It is important to investigate the causes of such inconsistent results before drawing any conclusions in such cases.} observed in the results of the review?}}} \newline
We identified no need for adaptation to SE in this item.\newline

\noindent\textbf{Item 15. \textit{\enquote{If they performed quantitative synthesis did the review authors carry out an adequate investigation of publication bias (small study bias) and discuss its likely impact on the results of the review?}}}
\begin{enumerate}[a)]
    \item The item is limited to quantitative synthesis. In SE, qualitative synthesis is used more frequently in SLRs. Discussing publication bias and its impact on the results is important, regardless of the type of synthesis performed, quantitative or qualitative.
    \item The response option includes a requirement to carry out graphical or statistical tests as well. The main aspect to cover in this item should be to check if the authors of the review have discussed publication bias and discussed its potential impact on review results.
\end{enumerate}
\noindent \underline{\textit{Changes:}} We introduced the following changes:
\begin{enumerate}[a)]
    \item The item is made more general by removing the word quantitative while also adapting its formulation for SE.
    \item The response option is also revised accordingly, i.e., removing the reference to the graphical or statistical tests. The revised response option aims to check if the publication bias and its impact on the results are discussed or not. \newline
\end{enumerate}

\noindent\textbf{Item 16. \textit{\enquote{Did the review authors report any potential sources of conflict of interest, including any funding they received for conducting the review?}}} \newline
We identified no need for adaptation to SE in this item.\newline

We call the resulting instrument that systematically adapts AMSTAR~2 for SE and supplements it SE guidelines and evidence QAISER~V0. This version was used in Step 4 (see Section~\ref{sec:method} for details of the process) for further validation.

\subsection{Changes in QAISER during the validation step}

This section presents the changes made in QAISER~V0 based on the feedback collected during the pilot and external validation steps.

Besides several editorial changes, the pilot validation resulted in the following two main changes in QAISER~V0:

\begin{enumerate}[1)]
    \item Addition of a new item on the need for undertaking the review (see QAISER Item 1 in Appendix A): In QAISER~V0, establishing the need for undertaking a review was listed as one of the response options to score 'Partial Yes' under Item 1. During the discussions in the pilot validation, we agreed with the senior colleague to give more importance to establishing the need for the review step. The number of SLRs performed in SE is increasing every year. At times, there are multiple SLRs on the same topic. Thus, there is a need to establish if it is relevant to undertake a new SLR on a topic~\cite{MendesWFK20,kitchenham2015evidence,PETERSEN20151}. The authors of the review should justify the need for undertaking the review. To score 'Yes' on this new item in QAISER, the review should have 1) discussed the related existing reviews (if any) and established the need for another review by highlighting the gap, or 2) established the need to aggregate the evidence on a topic, if there exist no reviews on the topic.
    \item Addition of a new response option under the synthesis related item in QAISER~V0 (Item 11): Agreeing with the suggestion of the senior colleague, we added another response options under Item 11 in QAISER to check how effectively the authors of the review have linked the answers and interpretations with the data extracted from the primary studies. The new response option is described as: \textit{\enquote{Provided a clear trace linking the answers of review questions and interpretations to the data from the included primary studies}}.
\end{enumerate}

The revised QAISER, after the pilot validation step, was shared with the external experts for further validation. The external experts provided several improvement suggestions. We provide a summary of the main suggestions related to items and response options in the following: 

\begin{itemize}
    \item Introduce an item about recommendations: SLRs are supposed to provide evidence-based input to practitioners and researchers to aid them in making informed decisions. QAISER did not have any item that specifically covered this aspect. The external experts suggested including an item that checks if the review provides appropriate recommendations and conclusions based on the review results. Agreeing with the external reviewer's suggestion, we added a new item about recommendations and conclusions in QAISER (see QAISER Item 14 in Appendix A).
    
    \item Remove the item about sources of funding (see AMSTAR~2 Item 10 described in Section~\ref{subsec:QAISERV0}): The item deals with the reporting of the sources of funding for the included studies. The external experts suggested to remove it as they did not find it relevant in SE context. We removed this item from QAISER. 
    
    \item Introduce 'Partial Yes' scale: Some items (Items 1, 5, 6, and 10) had a binary Yes/No scale. The external experts suggested introducing a third scale value of 'Partial Yes' to make them more flexible. We introduced a 'Partial Yes' option under these items and included the minimum acceptable requirements as response options (see QAISER Items 1, 5, 6, and 10 in Appendix A).
    
    \item Quality focus:  Assessing SLRs is not only about the presence or absence of an aspect; it is largely a subjective judgment concerning decisions and measures taken by the authors. To incorporate this suggestion, we introduced adjectives such as adequately, reliable, and appropriate in several items to assess SLRs' subjective nature better.
   
    \item Modifications to the protocol-related item (see AMSTAR Item 2 described in Section~\ref{subsec:QAISERV0}): The external experts suggested simplifying the response options for the 'Partial Yes' scale. We moved justification of any deviations from the protocol from 'Partial Yes' to the 'Yes' scale. Furthermore, threats to validity and details of conducting the review were removed from the 'Partial Yes' scale. We also removed a response option about heterogeneity from the 'Yes' scale. It was not deemed a necessary part of a protocol by the experts (see the revised description of QAISER Item 2 in Appendix A).
    \item Modifications to the heterogeneity-related item (see AMSTAR Item 14 described in Section~\ref{subsec:QAISERV0}): The external experts did not find this item to be essential for the systematic reviews in software engineering. The item is more relevant for meta-analysis studies, which are not common in software engineering. We replaced the heterogeneity concept with the characteristics of the primary studies. Some differences in the results of the primary studies may be due to the variations in the studies' characteristics, e.g. if the participants in different studies are students or practitioners. Therefore, in the case when there are differences in the results of the primary studies, the authors of the review should perform an analysis to see if the differences are due to the variations in the primary studies' characteristics.
\end{itemize}

\section{Results}
\label{sec:results}

In this section, we present the main outcomes of our study, i.e., QAISER after the validation step. We also describe how QAISER can be used to appraise an SLR. 

QAISER aims to support appraisers of SLRs in SE by raising important questions about the reliability and the relevance of an SLR. Furthermore, by providing evidence-based and accepted best practices in software engineering research (i.e., established expectations in the SE field of a high quality SLR), it supports the judgement of the conformance and the likely impact of non-conformance on the reliability and relevance of an SLR.

The quality aspects of concern and related criteria in QAISER are based on available evidence and recommendations in the SE literature. Therefore, the availability of evidence and the specificity of guidelines is also reflected in the criteria used in QAISER. Thus, the responses in the instrument range from specific/concrete actions to broader/general suggestions/guidelines. QAISER supports appraisers in making a judgement about the overall reliability and relevance of an SLR.

QAISER has three levels of judgement: item level, group level and SLR level. The three levels are described in the following subsections. It should be noted that AMSTAR~2 does not include these levels. The levels are introduced to support the appraiser in moving towards an overall assessment of an SLR. Table~\ref{tab:QAISERinpaper} presents the groups and the items of QAISER, while the complete instrument is presented in Appendix A (see supplemental material). 

\begin{table*}[ht]
\footnotesize
\caption{QAISER items and groups.}
\label{tab:QAISERinpaper}
\begin{tabular}{p{2cm}p{14cm}}
\toprule
\textbf{Group} &
  \textbf{Item description and the relevant sources/references}\\
  \midrule
1. Motivation                      & Item 1: Did the authors of the review adequately justify the need for undertaking the review?~\cite{kitchenham2015evidence,PETERSEN20151,AMPATZOGLOU2019201}\\
\midrule
2. Plan                            & Item 2: Did the authors of the review establish a protocol prior to the conduct of the review?~\cite{Sheaj4008,AMPATZOGLOU2019201,kitchenham2015evidence} \\
\midrule
3. Identification and selection &
  \begin{tabular}[c]{@{}p{14cm}}Item 4: Did the authors of the review use a comprehensive literature search strategy?~\cite{Sheaj4008,kitchenham2015evidence,AMPATZOGLOU2019201,DARE} \\ Item 3: Are the review's inclusion and exclusion criteria appropriate?~\cite{DARE,kitchenham2015evidence,AMPATZOGLOU2019201} \\ Item 5: Did the authors of the review use a reliable study selection process?~\cite{Sheaj4008,kitchenham2015evidence,PETERSEN20151}\\ Item 7: Did the authors of the review provide a list of excluded studies, along with the justifications for exclusion, that were read in full text?~\cite{Sheaj4008,kitchenham2015evidence} \end{tabular}
  \\
  \midrule
4. Data collection and appraisal &
  \begin{tabular}[c]{@{}p{14cm}}Item 6: Did the authors of the review use a reliable data extraction process?~\cite{Sheaj4008,kitchenham2015evidence,AMPATZOGLOU2019201,PETERSEN20151}  \\ Item 8: Did the authors of the review provide sufficient primary studies' characteristics to interpret the results?~\cite{Sheaj4008,kitchenham2015evidence,AMPATZOGLOU2019201,DARE}\\ Item 9: Did the authors of the review use an appropriate instrument for assessing the quality of primary studies that were included in the review?~\cite{Sheaj4008,kitchenham2015evidence}\\ Item 10: Did the authors of the review use a reliable quality assessment process?~\cite{kitchenham2015evidence}\end{tabular}
  \\
  \midrule
5. Synthesis &
  \begin{tabular}[c]{@{}p{14cm}}Item 11: Were the primary studies appropriately synthesized?~\cite{DARE,kitchenham2015evidence,AMPATZOGLOU2019201,BudgenBDW18}\\ Item 12: Did the authors of the review account for quality of individual studies when interpreting/discussing the results of the review?~\cite{Sheaj4008,kitchenham2015evidence,AMPATZOGLOU2019201} \\ Item 13: Did the authors of the review account for primary studies' characteristics when interpreting/discussing the results of the review?~\cite{Sheaj4008,kitchenham2015evidence}  \end{tabular}
  \\
  \midrule
6. Recommendations and conclusions & Item 14: Did the authors of the review provide appropriate recommendations and conclusions from the review?~\cite{BudgenBDW18}   \\
\midrule
7. Conflict of interest            & Item 15: Did the authors of the review report their own potential sources of conflict of interest, including any funding they received for conducting the review?~\cite{Sheaj4008,kitchenham2015evidence}   \\
\bottomrule
\end{tabular}%
\end{table*}

\begin{table*}[ht]
\centering
\footnotesize
\caption{QAISER: group level assessment.}
\label{tab:groups}
\begin{tabular}{p{3cm}p{5cm}p{1.5cm}p{1.3cm}}
\toprule
\textbf{Group} &
  \textbf{Item ranking (Yes/ Partial Yes /No)} &
  \textbf{Impact} &
  \textbf{Comments} \\
  \midrule
1. Motivation                      & Item 1 (need):             &  &  \\
\midrule
2. Plan                            & Item 2 (protocol):         &  &  \\
\midrule
3. Identification and selection &
  \begin{tabular}[c]{@{}l@{}}Item 4 (search):\\ Item 3 (selection criteria):\\ Item 5 (selection process):\\ Item 7 (excluded studies):\end{tabular} &
  &
  \\
  \midrule
4. Data collect-\newline ion and app-\newline raisal &
  \begin{tabular}[c]{@{}l@{}}Item 6 (data extraction):  \\ Item 8 (study characteristics):\\ Item 9 (quality criteria):\\ Item 10 (quality assessment process):\end{tabular} &
  &
  \\
  \midrule
5. Synthesis &
  \begin{tabular}[c]{@{}l@{}}Item 11 (synthesis):\\ Item 12 (considered study quality)\\ Item 13 (considered study characteristics)\end{tabular} &
  &
  \\
  \midrule
6. Recommendations and conclusions & Item 14 (recommendation): &  &  \\
\midrule
7. Conflict of interest            & Item 15 (their own):       &  & \\
\bottomrule
\end{tabular}%
\end{table*}

\begin{table*}[ht]
\footnotesize
\caption{Judging the confidence in the results of an SLR.}
\label{tab:Reliability}
\begin{tabular}{p{2cm}p{0.8cm}p{2cm}p{2cm}p{1.5cm}p{2.5cm}p{2.5cm}}
\toprule
\textbf{Confidence}       & \multicolumn{6}{c}{\textbf{Groups}} \\ \midrule
 &
  2. Plan &
  3. Identification and selection &
  4. Data collection and appraisal &
  5. Synthesis &
  6. Recommendations and conclusions &
  7. Conflict of interest \\ \midrule
"critically low" & \multicolumn{6}{c}{-- major weaknesses in groups 3, 4 and 5 --}       \\
"low"            & \multicolumn{6}{c}{-- major weaknesses in at most two of the groups 3, 4, 5 along with major weaknesses in groups 2 and 6  --}       \\
"moderate"       & \multicolumn{6}{c}{-- no major weakness in groups 3, 4, 5 and 7, but a major weaknesses in groups 2 or 6 --}       \\
"high"           & \multicolumn{6}{c}{-- only minor weaknesses in at the most two of the groups 3, 4, 5 and only a few minor weaknesses in groups 2, 6 and 7 --}     \\ 
\bottomrule
\end{tabular}%
\end{table*}

\begin{table*}[ht]
\centering
\footnotesize
\caption{Judging the relevance of an SLR.}
\label{tab:Relevance}
\begin{tabular}{p{2cm}p{7.0cm}}
\toprule
\textbf{Relevance}        & {\textbf{Groups} (1. Motivation \& 6. Recommendations and conclusions)} \\ \midrule
                 
"critically low" & {-- major weaknesses in group 1 and 6 --}\\
"low"            & {-- major weakness in either group 1 or 6 --} \\
"moderate"       & {-- minor weaknesses in both groups 1 and 6 --}\\
"high"           & {-- only a minor weakness in group 6 --}\\ \bottomrule
\end{tabular}%
\end{table*}

\subsection{QAISER: item level assessment}

The first level comprises 15 items formulated as questions. These questions are ordered to reflect the sequence of phases in the design, conduct, and reporting of a typical SLR. The criteria to meet the questions on the item level are stated in the form of acceptable responses for each of the questions. All items are evaluated on a scale with two values (Yes/No) or three values (Yes/Partial Yes/No), i.e., an assessment of the extent to which an SLR under review fulfils the stated criteria.

Each item in QAISER is formulated with the objective that it is self-contained and self-explanatory. However, there is an accompanying guidance document (Appendix B in the supplemental material) with a more detailed description of the items and their responses. We recommend that before applying QAISER, the guidance document should be read for at least the first time.

\subsection{QAISER: group level assessment} The external experts also provided a suggestion about clarifying the flow and sequence of the items in QAISER. To make the flow of the items more explicit and understandable, and to aggregate individual items into a logical cluster, we organized the 15 QAISER items into seven groups corresponding to the process and outcome of an SLR (see the first column in Table~\ref{tab:QAISERinpaper}): (1) motivation to conduct a review, (2) plan and its validation, (3) identification and selection, (4) data collection and quality appraisal, (5) synthesis, (6) recommendations and conclusions, and (7) conflict of interest.

At the group level, the assessment results on the item level are used as indicators for major and minor weaknesses based on their impact on the reliability and relevance of an SLR. Having completed the assessment of individual QAISER items, an appraiser should reflect on the impact of the weaknesses on the reliability and relevance of the SLR at the group level. Groups 1, 2, 6, and 7 consist of single items only, and are therefore relatively simple to reflect upon. A ''No'' rating on the corresponding items of these four groups indicates a major weakness at the group level. Groups 3, 4, and 5 consist of multiple items and are more complex to reflect upon. The appraisers should make an overall assessment after considering the ratings of all items in the groups. As a rule of thumb, we recommend that all items receiving a ``No'' should be considered as hinting at a major weakness in the group being assessed.

\subsection{QAISER: SLR level assessment} 

By progressively building on the first two levels, an appraiser judges the overall reliability and relevance of an SLR at the SLR level. Thus, considering the impact of weaknesses in related groups, i.e., relevance (mainly two groups: motivation, and recommendations and conclusions) and reliability (mainly the following five groups: plan, identification and selection, data collection and appraisal, synthesis, and conflict of interest). 

Once the reflection on the group level is complete, an appraiser should use it to provide the assessment about the two aspects at the SLR level as follows:

\begin{enumerate}[1)]
    \item Reliability of an SLR: The reliability of an SLR is assessed by rating the overall confidence in the results of an SLR as: high, moderate, low or critically low. Apart from group 1, all other groups are relevant while considering the confidence in the results of an SLR. As a rule of thumb, we recommend that the confidence in the SLRs with major weaknesses in groups 3, 4, and 5 should be considered ``critically low''.

    Table~\ref{tab:Reliability} provides guidance for interpreting weaknesses observed at the group level to select a confidence rating at the SLR level.
    
    \item Relevance of an SLR: The relevance of an SLR is also rated as high, moderate, low or critically low. Groups 1 and 6 are considered when making a judgement about the relevance of an SLR. As a rule of thumb, we recommend that the relevance of an SLR be judged to be ``critically low'' if there are major weaknesses in both groups 1 and 6. Table~\ref{tab:Relevance} provides guidance in selecting a relevance rating based on the weaknesses in groups 1 and 6.
\end{enumerate}

The appraisers can and should decide which groups of items are more/less critical for the specific review being assessed. The guidance provided in Tables~\ref{tab:Reliability} and~\ref{tab:Relevance} is one such recommendation for using group-level assessment in Table~\ref{tab:groups} for assessing the reliability and relevance of an SLR.

\section{Reliability of QAISER}
\label{sec:reliability}
In this section, we highlight three aspects that contribute to the reliability of QAISER as a potentially effective instrument for assessing the quality of SLRs in SE.

\begin{enumerate}[1)]
    \item The relevance of AMSTAR and AMSTAR 2 validations: The original AMSTAR~\cite{shea2007development} consisted of 11 appraisal items. Based on community feedback, AMSTAR 2 was proposed consisting of 16 items with an increased focus on the risk of bias evaluation and the possibility to assess SLRs that may have non-randomized studies. Both AMSTAR and AMSTAR 2 have been used and validated extensively (cf.\cite{gates2018evaluation,pieper2015systematic,shea2007external}). These validation efforts provide credibility to QAISER as well, as most of its items (12 out of 15) are adapted from AMSTAR 2.
    
    \item Comparison with DARE: DARE \cite{DARE} is the most frequently used criteria to assess the quality of SLRs. Several essential aspects related to the quality of SLRs are not covered in DARE, e.g., justifying the need to conduct a review, establishing a protocol prior to performing the review, study selection process, data extraction process, and quality assessment process. Furthermore, three of the DARE criteria (including the important criterion about synthesis) are limited to checking the presence/absence of different aspects, rather than their appropriateness, e.g., \textit{if the inclusion and exclusion criteria are reported or not?} QAISER not only covers aspects that are missing in DARE, but it also focuses on quality aspects of different criteria—for example, checking the appropriateness of inclusion and exclusion criteria rather than only focusing on the mere reporting of such criteria in the review report.
    
    \item External validation. We followed a systematic process (see Section \ref{sec:method} for details) to adapt AMSTAR 2 for SE and to introduce additional aspects based on the recommendations in the systematically identified EBSE literature. Four leading experts then reviewed the proposed instrument to check the appropriateness, completeness, and understanding of its items (refer to Section \ref{sec:method} for details about the validation step). The experts recommended some changes, which we incorporated in the revised version of QAISER. The experts did not suggest any further changes in the revised version.
\end{enumerate}

\section{Support for Applying QAISER}
\label{sec:application}
In line with AMSTAR 2, we also developed a guidance document for supporting appraisers in applying QAISER. The guidance document describes the following aspects for each QAISER item:

\begin{enumerate}[1)]
    \item What is the item about? We provide a brief description of the item.
    \item How to assess the item?  We explain what an appraiser needs to check for assigning 'Partial yes', 'Yes', and 'No' ratings.
    \item Where to find the relevant information to assess the item? We provide hints concerning which sections of the review papers are most likely to have the information needed to assess the item.
\end{enumerate}

To further support the application of QAISER, we developed a spreadsheet that operationalizes the QAISER instrument. The first author used the spreadsheet to apply the QAISER instrument on three SLRs that the second author selected. The selected SLRs were all previously assessed using DARE criteria with the highest rating. The purpose of selecting only the high-ranking SLRs on DARE criteria was to illustrate the usefulness of QAISER in supporting appraisers in performing a more fine-grained and thorough critical appraisal compared to DARE. The guidance document, QAISER instrument, and the spreadsheet corresponding to the three example applications are all available online\footnote{\url{https://drive.google.com/drive/folders/1p7OUEfqQTF4dY3e_OX_OHiyi_tC4E_cU?usp=sharing}}. Researchers could look at the three examples as additional support complementing the guidance document.

QAISER supported us in identifying additional weaknesses in all three SLRs that scored high on the DARE criteria. The three SLRs have different weaknesses and, taken together, include the following aspects: study selection process, data extraction process, quality assessment instrument and process, and conflicts of interest. Furthermore, for those aspects that are already covered in DARE (e.g., inclusion and exclusion criteria and search strategy), we were able to perform a more fine-grained appraisal. It resulted in identifying some additional weaknesses, e.g., appropriate synthesis and comprehensiveness of the search strategy. For details, we suggest the interested readers refer to the online spreadsheet. A unique characteristic of QAISER, which is not available in DARE, is how it allows appraisers to progressively arrive at the SLR level assessment from the group and item level assessments. In the selected SLRs, this mechanism helped us first combine the item level ratings to identify the weaknesses at the group level and then use the identified group level weaknesses to arrive at the overall judgment at the SLR level. Finally, we also found the option to add comments extremely helpful in reflecting on our ratings and corresponding justifications.

\section{Threats to validity}
\label{sec:ValidityThreats}

In this research, we aimed to propose an instrument for appraising systematic literature reviews in software engineering. In the design and conduct of this research, the following two objectives guided us:\newline
\begin{enumerate}[1)]
    \item To develop an instrument that is comprehensive, practical and appropriate for software engineering. Thus, the instrument shall cover all essential elements concerning the quality of an SLR, assist the appraiser when judging the quality of an SLR, and take into account the SE body of knowledge.
    \item To reduce the researcher's bias in the development of the instrument.
\end{enumerate}

The two main threats to validity identified concerned researcher bias and applicability of QAISER. The researchers come from the same affiliation, which creates a risk of having a coherent view on research. When creating an instrument for use by the research community, there is a risk that the instrument is hard to understand, and hence limiting its applicability.

To achieve the two objectives above and mitigate the threats to validity, we undertook the following actions:   
\begin{itemize}
\item \textbf{Use of an established tool as a foundation:} We used AMSTAR~2 as the starting point for our work as it is a well used and validated tool~\cite{ALI201948}.

\item \textbf{A systematic and rigorous process:} As described in Section~\ref{sec:method}, we followed a systematic approach for the development of QAISER. All data collection, analysis and interpretation involved at least two researchers. A third researcher independently reviewed the outcomes from several individual phases in the study. We maintain traceability for the adaptations in the existing tool by documenting the reasons, sources consulted and the changes. 

\item \textbf{Validation with leading experts in the field:} The experts consulted to validate QAISER include some of main contributors of the methodological guidelines for designing, conducting and reporting secondary studies in SE. They have also authored several SLRs and have conducted several studies reporting critical evaluations of existing SLRs in SE.

\item \textbf{Comprehensive coverage of SE body of knowledge:} We used a systematic approach to identify and select a representative list of sources to capture the current best practices and evidence for inclusion in QAISER.
\end{itemize}

With these actions, we have tried to mitigate the major threats to the validity of our research. However, a missing aspect in our research is the evaluation of the instrument in use (focusing on aspects like usability and reliability of QAISER). By making QAISER and guidance for its usage publicly available, we hope that we and others in the field will address this limitation in the future. 

\section{Discussion}
\label{sec:discussion}
Given the lack of an appraisal instrument for assessing the quality of SLRs in SE, we developed QAISER. As presented in the introduction (see Section~\ref{sec:intro}), researchers in SE have used the criteria in DARE~\cite{DARE} for assessing the quality of SLRs, although it comes with limitations~\cite{AliU18}. Furthermore, to simply use an appraisal instrument, such as AMSTAR~2, from another discipline also comes with issues as illustrated in the development of QAISER. There was a need to adapt AMSTAR~2 to SE, and hence AMSTAR~2 is not an option by itself. The differences between disciplines need to be captured in the appraisal instrument.
 
QAISER takes its starting point from a well-established appraisal instrument from another field, i.e., AMSTAR~2 from the field of evidence-based healthcare. Furthermore, QAISER incorporates best practices from the conduct of SLRs in SE. Thus, QAISER is well-grounded in the literature and contributes to taking the quality assessment of SLRs in SE one step forward.
 
 The objective of QAISER is to support appraisers of SLRs in SE. QAISER is not intended as an instrument for scoring the quality of SLRs. On the contrary, QAISER is intended to support appraisers by covering the most salient quality aspects of an SLR. The expertise of the individual appraisers is crucial, and it cannot be replaced with an appraisal instrument such as QAISER.
 
 Although the main objective is to support appraisers in assessing quality, we believe that authors of SLRs may also use QAISER to help them with improving the quality of their SLR before submitting the research for assessment. In the best of worlds, each submitted SLR is of high quality at the submission stage. It should be noted that the quality of an SLR is highly influenced by the quality of the primary studies included. The need to assess the quality of primary studies is highlighted by, for example, Dyb{\aa} and Dings{\o}yr~\cite{dyba_strength_2008}, and Yang et al.~\cite{YANG2020106397}. With the same objective, Wohlin highlights the need to write for synthesis when publishing primary studies~\cite{Wohlin14Synthesis}. 
 
We recommend all users of QAISER to not only look at the appraisal instrument itself but also the accompanying guidance document. The latter is particularly important when using the instrument for the first couple of times. We have also made available online a spreadsheet operationalizing the QAISER and three example assessments to further support appraisers in using QAISER.

The items and their response options in QAISER are intended to help highlight areas with weaknesses (or room for improvement). Given that assessment is prone to bias, we have deliberately chosen to have two or three levels for assessing each item. More levels may increase the risk for appraiser bias, although it may also benefit since the scale becomes more fine-grained. However, since QAISER is geared towards supporting appraisers of SLRs, we leave it to each appraiser to tune the feedback in writing using the comments option provided with each item, rather than having a more fine-grained scale.

When using QAISER for a mapping study, some items or questions may be less applicable than for an SLR, for example, the item concerning synthesis. We did consider adding an option of ``not applicable'' for mapping studies. We have chosen not to make the appraisal instrument more complex by adding the "not applicable" option. Thus, we leave it to each appraiser to decide if something is not applicable for a mapping study. Our preference is to leave freedom to the appraiser, given that SLRs and mapping studies may come in different shapes and colors. Assessing SLRs and mapping studies is a subjective endeavour, and the objective of any appraisal instrument should be to support the expert appraiser.

\section{Conclusion and Future Work}
\label{sec:conclusion}

QAISER, as an appraisal instrument for SLRs in SE, is built on a well-established appraisal instrument from another discipline (AMSTAR~2), and a set of guidelines, checklists, and experiences from SE. Furthermore, four external experts on SLRs in SE have reviewed an earlier version of QAISER, and QAISER has been revised based on their feedback. Thus, QAISER is well-founded, and hence it is ready for further validation through usage. 

QAISER includes 16 items and several response options for each item to assess for appraisers to arrive at an assessment for each item. QAISER provides support to consolidate the items on a group level, which is not done in AMSTAR~2. In QAISER, the items are consolidated into seven groups to support the appraiser to get a good overview of the strengths and potential weaknesses of an SLR. Moreover, QAISER has support for consolidating from the group level to the SLR level. The assessment of each group is systematically used to form an opinion about the overall quality of an SLR both in terms of reliability and relevance. AMSTAR~2 only provides an overall assessment of the confidence in the results. Given the importance of both reliability and relevance of the results for software engineering, we have provided support for both aspects.

In the future, we plan to evaluate the reliability and usability of QAISER by asking independent researchers to use it to assess the quality of selected SLRs. Based on such feedback, we plan to enhance QAISER further to support the software engineering community assessing SLRs.



\section*{Acknowledgements}
We would like to express our sincere thanks to the external experts: Prof. Daniela S. Cruzes, Prof. Barbara Kitchenham, Prof. Stephen G. MacDonell and Prof. Kai Petersen for their constructive feedback on QAISER.

We would also like to thank Prof. J\"urgen B\"orstler for his kind participation in the pilot of the study. His detailed feedback helped us to improve the planning and execution of the evaluations with external experts. We also extend our gratitude to Dr. Jefferson S. Moll{\'e}ri for providing the listing of articles from the work with CERSE.

This work has been supported by ELLIIT, a Strategic Area within IT and Mobile Communications, funded by the Swedish Government. The work has also been supported by research grants for the VITS project (reference number 20180127) and the OSIR project (reference number 20190081) from the Knowledge Foundation in Sweden.

\balance
\bibliographystyle{elsarticle-num}
\bibliography{bibliography}

\vspace{0.8cm}
\cleardoublepage
\noindent \textbf{Supplemental Material}: \newline
\onecolumn
\section*{Appendix A. QAISER instrument}
\label{AppendixA}

\begin{longtable}{p{7cm}p{8cm}p{1.9cm}}
	\caption{QAISER Instrument.}\\
	\label{tab:QAISER}
	 &  &  \\
	\endfirsthead
	\multicolumn{3}{c}%
	{{\bfseries Table \thetable\ continued from previous page}} \\
	 &  &  \\
	\endhead
	\toprule
	
	\multicolumn{3}{p{16.6cm}}{\textbf{1.     Did the authors of the review adequately justify the need for undertaking the review? \cite{kitchenham2015evidence,PETERSEN20151,AMPATZOGLOU2019201}}} \\
	\midrule
	For Partial Yes: & For Yes: &  \\
	\midrule
    The authors of the review should have: &  As for partial yes, plus the authors of the review should also have ALL of the following:\\
	\midrule
	{$\square$} Identified existing related reviews on the topic, or explained that no related review exists.\newline
	&
	{$\square$} Discussed related existing reviews on the topic, if any. \newline
	{$\square$} Established a scientific or practical need for their review \cite{kitchenham2015evidence}.
	\newline &
	{$\square$} Yes \newline
	{$\square$} Partial Yes\newline
	{$\square$} No \\
	\midrule
	Comments:&\\
	&&\\
	\midrule
	
	\multicolumn{3}{p{16.6cm}}{\textbf{2.     Did the authors of the review establish a protocol prior to the conduct of the review? \cite{Sheaj4008,AMPATZOGLOU2019201,kitchenham2015evidence}}} \\
	\midrule
	For Partial Yes: & For Yes: &  \\
	\midrule
    The authors of the review state that they had a written protocol that is publicly available that should specify ALL of the following: & As for partial yes, plus ALL of the following: &  \\
	\midrule
	{$\square$} Appropriate review questions \cite{Sheaj4008,kitchenham2015evidence,PETERSEN20151} \newline
	{$\square$} Search process \cite{Sheaj4008,kitchenham2015evidence} \newline
	{$\square$} Study selection process \cite{Sheaj4008,kitchenham2015evidence,AMPATZOGLOU2019201} \newline
	{$\square$} Data extraction process \cite{kitchenham2015evidence} \newline
	{$\square$} Study quality assessment process \cite{Sheaj4008,kitchenham2015evidence}\newline
	{$\square$} An outline of the data synthesis plan \cite{Sheaj4008,kitchenham2015evidence} &
	{$\square$} The justifications of any deviations from the protocol should be documented \cite{Sheaj4008,AMPATZOGLOU2019201,kitchenham2015evidence}. \newline
	{$\square$} The protocol should have been internally validated by piloting selection criteria, search strings, data extraction and synthesis processes  \cite{kitchenham2015evidence}. \newline
	{$\square$} The protocol should have been validated by an external reviewer \cite{kitchenham2015evidence}. \newline
	
	&
	{$\square$} Yes \newline
	{$\square$} Partial Yes\newline
	{$\square$} No \\
	\midrule
	Comments:&\\
	&&\\
	\midrule
	\multicolumn{3}{p{16.6cm}}{\textbf{3.     Are the review's inclusion and exclusion criteria appropriate?} } \\
	\midrule
	
	For Yes: the review should have ALL of the following: & &  \\
	\midrule
	\multicolumn{2}{p{16cm}}{{$\square$} Reported the inclusion and exclusion criteria \cite{DARE,kitchenham2015evidence,AMPATZOGLOU2019201}.
	\newline
	{$\square$} The criteria are aligned with the review questions. \newline
	{$\square$} Provided appropriate justifications for any restrictions used in the inclusion and exclusion criteria (e.g.,topic-related scoping restrictions, time-frame,  language, study type, and peer reviewed works only) \cite{kitchenham2015evidence,Sheaj4008}.} & 
	{$\square$} Yes \newline
	{$\square$} No \\
	&  &  \\
	\midrule
	Comments:&\\
	&& \\
	\midrule
	\multicolumn{3}{p{16.6cm}}{\textbf{4.     Did the authors of the review use a comprehensive literature search strategy? \cite{Sheaj4008,kitchenham2015evidence,AMPATZOGLOU2019201,DARE} }}  \\
	\midrule
	When database search is used as the main method &  &  \\
	\midrule
	For Partial Yes: & For Yes: &  \\
	\midrule
	The review should have All of the following: & As for partial yes, plus the review should also have used: &  \\
	\midrule
	{$\square$} An appropriate process for constructing the search strings including piloting \cite{kitchenham2015evidence,AMPATZOGLOU2019201} \newline
	{$\square$} Search process validation based on an appropriate level of recall and precision using a known-set of papers \cite{kitchenham2015evidence,AMPATZOGLOU2019201,PETERSEN20151}  \newline
	{$\square$} At least one relevant indexing database (e.g., Scopus) in combination with relevant publisher databases (e.g., IEEE and ACM) \cite{kitchenham2015evidence} \newline
	{$\square$} Appropriately documented the search process (e.g., known-set, search strings, and search results) \cite{kitchenham2015evidence,AMPATZOGLOU2019201, AliU18}  & 
	{$\square$} At least one additional search method (e.g., snowballing, manual search, or use DBLP or Google scholar of key researchers) \cite{Sheaj4008,kitchenham2015evidence,AMPATZOGLOU2019201} &
	 {$\square$} Yes \newline
	{$\square$} Partial Yes\newline
	{$\square$} No \\
	
	&  &  \\
	\midrule
	When snowballing is used as the main method &  &  \\
	\midrule
	For Partial Yes: & For Yes: &  \\
	\midrule
	The review should have ALL of the following: & As for partial yes, plus the review should also have used either ONE of the following: &  \\
	\midrule
	{$\square$} Appropriately justified the use of snowballing as the main method \cite{kitchenham2015evidence} \newline
	{$\square$} Selected an appropriate start/seed set \cite{Wohlin:2014:GSS:2601248.2601268} \newline
	{$\square$} Performed an acceptable number of backward and forward snowballing iterations \cite{Wohlin:2014:GSS:2601248.2601268} & 
	{$\square$} At least one additional search method (e.g., manual search, or use DBLP or Google scholar of key researchers) \cite{Wohlin:2014:GSS:2601248.2601268,kitchenham2015evidence} \newline
	{$\square$} Snowballing iterations until no new papers were found \cite{Wohlin:2014:GSS:2601248.2601268}&
	 {$\square$} Yes\newline
	 {$\square$} Partial Yes\newline
	 {$\square$} No \\
	 
	 \midrule
	Comments:&\\
	 &  &  \\
	\midrule
	
	\multicolumn{3}{p{16.6cm}}{\textbf{5.     Did the authors of the review use a reliable study selection process? \cite{Sheaj4008,kitchenham2015evidence,PETERSEN20151} }} \\
	\midrule
	
	For Partial Yes: & For Yes, either ONE of the following: &  \\
	\midrule
	{$\square$} At least two authors of the review selected a representative sample of eligible studies and achieved good agreement, with the remainder selected by one review author \cite{Sheaj4008, kitchenham2015evidence, AMPATZOGLOU2019201}. \newline 
	&
	{$\square$} At least two authors of the review independently agreed on selection of eligible studies, reached consensus on which studies to include, and reported the agreement level \cite{Sheaj4008,kitchenham2015evidence,PETERSEN20151}.\newline 
	{$\square$} OR if only a sample of studies were selected by two authors, all excluded studies were reviewed by at least two authors of the review.  \newline
	&
	{$\square$} Yes \newline
	{$\square$} Partial Yes\newline
	{$\square$} No \\
	
	\midrule
	Comments:&\\
	&&\\
	\midrule
	
	\multicolumn{3}{p{16.6cm}}{\textbf{6.     Did the authors of the review use a reliable data extraction process? \cite{Sheaj4008,kitchenham2015evidence,AMPATZOGLOU2019201,PETERSEN20151} }} \\
	\midrule
	
	For Partial Yes: & For Yes: &  \\
	\midrule
	{$\square$} At least two authors of the review extracted data from a sample of included studies and achieved good agreement, with the remainder extracted by one review author \cite{Sheaj4008,kitchenham2015evidence,PETERSEN20151} \newline
	&
	{$\square$} At least two authors of the review achieved consensus on which data to extract from the included studies \cite{Sheaj4008,kitchenham2015evidence,AMPATZOGLOU2019201,PETERSEN20151} &
	{$\square$} Yes \newline
	{$\square$} Partial Yes\newline
	{$\square$} No \\
	
	\midrule
	Comments:&\\
	&&\\
	\midrule
	
	\multicolumn{3}{p{16.6cm}}{\textbf{7.     Did the authors of the review provide a list of excluded studies, along with the justifications for exclusion, that were read in full text? \cite{Sheaj4008,kitchenham2015evidence} }}\\
	\midrule
	For Partial Yes, the review should have: & For Yes, the review should also have: &  \\
	\midrule
	{$\square$}  Provided a list of all potentially relevant studies that were read in full text, but excluded from the review \cite{Sheaj4008,kitchenham2015evidence} & 
	{$\square$}  Justified the exclusion from the review of each potentially relevant study that was read in full text \cite{Sheaj4008} & 
	{$\square$} Yes\newline
	{$\square$} Partial Yes\newline
	{$\square$} No \\
	
	\midrule
	Comments:&\\
	&  &  \\
	\midrule
	\multicolumn{3}{p{16.6cm}}{\textbf{8.  Did the authors of the review provide sufficient primary studies' characteristics to interpret the results? \cite{Sheaj4008,kitchenham2015evidence,AMPATZOGLOU2019201,DARE} }}  \\
	\midrule
	For Yes, the review should have described ALL of the following: &  &  \\
	\midrule
	\multicolumn{2}{p{16cm}}{{{$\square$} Populations \cite{Sheaj4008}\newline	
	{$\square$} Interventions, when applicable \cite{Sheaj4008} \newline
	{$\square$} Outcomes, when applicable \cite{Sheaj4008}\newline
	{$\square$} Study types \cite{Sheaj4008} \newline
	{$\square$} Study contexts \cite{Sheaj4008}}} & 
	{$\square$} Yes\newline{$\square$} No \\
	
	\midrule
	Comments:&\\
	&  &  \\
	\midrule
	\multicolumn{3}{p{16.6cm}}{\textbf{9.     Did the authors of the review use an appropriate instrument for assessing the quality of primary studies that were included in the review? \cite{Sheaj4008,kitchenham2015evidence} }} \\
	\midrule
	\multicolumn{2}{p{16cm}}{For Yes, the review should have used appropriate instruments for different types of studies included in the review. An appropriate instrument would have questions related to ALL of the following \cite{kitchenham2015evidence}: }  &  \\
	\midrule
	\multicolumn{2}{p{16cm}}{{{$\square$} \textit{"The goals, research questions, hypotheses and outcome measures"} \cite{kitchenham2015evidence}\newline 
	{$\square$} \textit{"The study design and the extent to which it is appropriate to the study type"} \cite{kitchenham2015evidence} \newline
	{$\square$} \textit{"Study data collection and analysis and the extent to which they are appropriate given the study design"} \cite{kitchenham2015evidence} \newline
	{$\square$} \textit{"Study findings, the strength of evidence supporting those findings, the extent to which the findings answer the research questions, and their value to practitioners and researchers"} \cite{kitchenham2015evidence} }}  & 
	{$\square$} Yes\newline
	{$\square$} No \\
			
	\midrule
	Comments:&\\
	&  &  \\
	\midrule
	
	\multicolumn{3}{p{16.6cm}}{\textbf{10.     Did the authors of the review use a reliable quality assessment process? \cite{kitchenham2015evidence} }} \\
	\midrule
	
	For Partial Yes: & For Yes: &  \\
	\midrule
	{$\square$} At least two authors of the review performed quality assessment of a representative sample of eligible studies and achieved good agreement, with the remainder performed by one review author \cite{kitchenham2015evidence} \newline
	&
	{$\square$} At least two authors of the review independently performed quality assessment of eligible studies, reached consensus, and reported the agreement level \cite{kitchenham2015evidence} &
	{$\square$} Yes \newline
	{$\square$} Partial Yes\newline
	{$\square$} No \\
	
	\midrule
	Comments:&\\
	&&\\
	\midrule
	
	\multicolumn{3}{p{18cm}}{\textbf{11.  Were the primary studies appropriately synthesized? \cite{DARE,kitchenham2015evidence,AMPATZOGLOU2019201,BudgenBDW18}}} \\
	\midrule
	For Yes, the review should have ALL of the following: &  &  \\
	\midrule
	\multicolumn{2}{p{16cm}}{{$\square$}   Selected an appropriate synthesis method given the review questions and extracted data \cite{kitchenham2015evidence,AMPATZOGLOU2019201,BudgenBDW18}\newline
	{$\square$} Applied the selected synthesis method appropriately \newline
	{$\square$} Provided a clear trace linking the answers of review questions and interpretations to the data from the primary studies } & 
	{$\square$} Yes\newline
	{$\square$} No \\
	
	\midrule
	Comments:&\\
	&&\\
	\midrule
	\multicolumn{3}{p{18cm}}{\textbf{12.  Did the authors of the review account for quality of individual studies when interpreting/discussing the results of the review? \cite{Sheaj4008,kitchenham2015evidence,AMPATZOGLOU2019201} }}  \\
	\midrule
	For Yes, either ONE of the following: &  &  \\
	\midrule
	\multicolumn{2}{p{16cm}}{{$\square$}  Included only high-quality studies \cite{Sheaj4008,kitchenham2015evidence} \newline
	{$\square$}  OR, the authors have categorized the analysis and interpretation of results based on study quality \cite{Sheaj4008,kitchenham2015evidence} }  & 
	{$\square$} Yes\newline
	{$\square$} No \\
	
	\midrule
	Comments:&\\
	&&\\
	\midrule
	
	\multicolumn{3}{p{18cm}}{\textbf{13.  Did the authors of the review account for primary studies' characteristics when interpreting/discussing the results of the review?}}  \\
	\midrule
	For Yes, either ONE of the following: &  &  \\
	\midrule
	\multicolumn{2}{p{16cm}}{{$\square$}  There were no significant similarities or differences to warrant a separate analysis \newline
	{$\square$}  OR, the authors have appropriately accounted for study characteristics and discussed their impact on the review results \cite{Sheaj4008,kitchenham2015evidence} }  & 
	{$\square$} Yes\newline
	{$\square$} No \\
	
	&&\\
	\midrule

	\multicolumn{3}{p{16.6cm}}{\textbf{14.     Did the authors of the review provide appropriate recommendations and conclusions from the review? \cite{BudgenBDW18} }} \\
	\midrule
	
	For Partial Yes, the review should have: & For Yes, the recommendations and conclusions should also be &  \\
	\midrule
	{$\square$} Provided satisfactory recommendations and conclusions based on the review results \newline
	&
	{$\square$} Clearly traceable back to the review results \newline
	{$\square$} Clearly targeting specific stakeholders \newline
	{$\square$} Well aligned with the upfront motivation for undertaking the review, or are any deviations well explained \newline
	{$\square$} Providing new valuable insights to the community \newline
	&
	{$\square$} Yes \newline
	{$\square$} Partial Yes\newline
	{$\square$} No \\
	
	\midrule
	Comments:&\\
	&&\\
	\midrule

	\multicolumn{3}{p{18cm}}{\textbf{15.  Did the authors of the review report their own potential sources of conflict of interest, including any funding they received for conducting the review? \cite{Sheaj4008,kitchenham2015evidence}}}  \\
	\midrule
	For Yes, either ONE of the following: &  &  \\
	\midrule
	\multicolumn{2}{p{16cm}}{{$\square$}   The authors reported no competing interests \newline
	{$\square$}  OR, the authors described their funding sources \cite{Sheaj4008,kitchenham2015evidence} and how they managed potential conflicts of interest \cite{Sheaj4008}}  &
	{$\square$} Yes\newline
	{$\square$} No \\
	\midrule
	Comments:&\\
	\bottomrule
\end{longtable}

\newpage\clearpage
\onecolumn
\section*{Appendix B. QAISER Guidance Document}
\label{AppendixB}
In this document, we provide further guidance to support a consistent interpretation of items in QAISER.
\newline 

\noindent \textbf{Item 1: Did the authors of the review adequately justify the need for undertaking the review?}
\newline
A large number of SLRs are reported in software engineering every year. A review should be initiated on the basis of a practical or scientific need. The authors of the review should also extensively search for any existing reviews or mapping studies on the topic. The authors should only continue planning the review if there are no existing ones that are up to date on the specific area \cite{kitchenham2015evidence}. Mendes et al. \cite{MendesWFK20} provide support to decide if a review should be updated.

To score 'Partial Yes', appraisers should check if the authors of the view have made a sufficiently extensive effort to identify related reviews on the topic.  

To score 'Yes', appraisers should ensure that the authors of the review have established the need for undertaking the review. If there are existing reviews on the same topic, the authors need to establish the need by highlighting the gap in the existing reviews, and explaining how their review is going to fill the gap. In case there are no existing reviews on the topic, the authors explain why is it essential to aggregate the evidence on the topic.

The information about the need for review is typically described in the background or related work sections of the report.
\newline

\noindent \textbf{Item 2:	Did the authors of the review establish a protocol prior to the conduct of the review?}
To reduce the risk of bias, it is important that the authors of the review have developed and validated a written protocol before commencing the review. 

To score 'Partial Yes', appraisers should first check that the protocol is accessible and the review report describes where and how can it be accessed. Furthermore, the protocol should have documented appropriate review questions, the processes for search, study selection, data extraction, quality assessment and at least an outline for the data synthesis plan. 

To rate 'Yes', the protocol should have been validated both internally and by an independent reviewer and the authors of the review should have clearly documented and justified any deviations from the protocol and discuss their impact on the study. If the appraisers notice that the review report contains unexplained deviations from the protocol, they should downgrade the rating. 

The above information about the protocol is typically described in the methodology section of the review report. \newline

\noindent \textbf{Item 3:	Are the review's inclusion and exclusion criteria appropriate?}
A review should use documented selection criteria \cite{AMPATZOGLOU2019201,Sheaj4008,kitchenham2015evidence,PETERSEN20151,DARE}.

To score 'Yes', appraisers should ensure that the authors of the review have justified any restrictions, e.g., on research designs, the time frame of publication, and the type of publications imposed in the selection process. Furthermore, the justification should also address the likely impact of the restrictions on the studied population and the generalization of the findings. 

The selection criteria and the justifications for any restrictions are expected to be found in the methodology or limitations/threats to the validity section of the review report. Furthermore, some of the exclusion criteria may have been implemented in the search process.
\newline

\noindent \textbf{Item 4:	Did the authors of the review use a comprehensive literature search strategy?}
A comprehensive search strategy is important to maximize the coverage of the relevant literature. The authors of the review should provide a justification for using a particular search method (e.g., database or indexing service search or snowballing) as a primary method for searching the relevant literature. 

To rate 'Partial Yes', appraisers should check the following in case of a database or indexing service search as the primary search method:
\begin{itemize}
    \item The authors of the review have validated their search process by comparing their search results with a known-set of papers. The known-set of papers are the relevant papers that are already known to the authors of the review based on, for example, manual search or their knowledge of the review topic. The validation is performed by computing recall and precision using the search results and the known set of relevant papers (for details, refer to \cite{kitchenham2015evidence}).
    \item The authors of the review have used an appropriate process for identifying the search terms, synonyms and constructing the search strings.
    \item The authors of the review have used a combination of publisher databases and indexing services. IEEE and ACM are the most relevant publisher databases in software engineering, as they publish the most important and relevant conferences and journals in software engineering \cite{kitchenham2015evidence}. As a minimum, the authors should have used IEEE and ACM among the publisher databases and one indexing service (e.g. Scopus).
    \item The authors of the review have documented the search process. Appropriate documentation of the search process is important to ensure repeatability and transparency. The authors of the review should document: general and database specific search strings, total and database specific search results, search filters (e.g., years) used, date when the search strings were applied, known-set of papers used for validation, and validation measures (i.e. recall and precision) for details see \cite{AliU18}. 
    \end{itemize}
To rate 'Yes', the authors of the review should have also used at least one additional search method (e.g., snowballing).

To rate 'Partial Yes', appraisers should check the following in case of a snowballing search as the primary search method:
\begin{itemize}
     \item The authors of the review have used an appropriate process for identifying the seed set for starting the snowballing procedure. The way of identifying the seed set is well-documented and -motivated.
     \item The authors have iterated the snowballing procedure until no more papers are found.
     \item The authors of the review have validated their search process by comparing their search results with a known-set of papers. The known-set of papers are the relevant papers that are already known to the authors of the review based on, for example, manual search or their knowledge of the review topic. The validation is performed by computing recall and precision using the search results and the known set of relevant papers (for details, refer to \cite{kitchenham2015evidence}).
    \item The authors of the review have documented the search process. Appropriate documentation of the search process is important to ensure repeatability and transparency. The authors of the review should document: identification of the seed set, the different iterations conducted, known-set of papers used for validation, and validation measures (i.e., recall and precision). 
    \end{itemize}
To rate 'Yes', the authors of the review should have also used at least one additional search method (e.g., manual search of key journals or conference proceedings, or use DBLP or Google scholar of key researchers) or continuing snowballing iterations until no new papers are found).
\newline

\noindent \textbf{Item 5. Did the authors of the review use a reliable study selection process? }
To reduce bias and the possibility of making mistakes, the crucial step of inclusion and exclusion of the papers should involve at least two reviewers~\cite{AliP14,Sheaj4008}.

To rate 'Partial Yes', appraisers should check if at least two authors of the review selected a representative sample of eligible studies and achieved good agreement, with the remainder selected by one review author. A single reviewer should only proceed with the selection after a Kappa score indicating strong agreement between multiple authors of the review has been reached.

To rate 'Yes', appraisers should check that one of the two following processes are followed during study selection: 1) two authors of the review independently performed study selection on all eligible studies, reached consensus and also reported the agreement level, 2) two authors of the review selected a sample of eligible studies, achieved good agreement level, with the remainder selected by one review author, but in that case all excluded studies must be reviewed by at least two authors of the review. A single reviewer should only proceed with the selection after a Kappa score indicating strong agreement between multiple authors of the review has been reached. However, even in this case, the excluded studies should be reviewed by other authors of the review as well. Appraisers should also check that the rules for inclusion and exclusion, and how these rules were applied and how any differences between reviewers were resolved are described. Furthermore, the report should also report the number of papers remaining at each stage~\cite{BudgenBDW18}.

The information about the study selection process is expected to be described in the methodology and results sections of the review.
\newline

\noindent \textbf{Item 6.	Did the authors of the review use a reliable data extraction process?}
To ensure repeatability of the study and to avoid bias, it is important that the data extraction is not solely performed by a single researcher.

To rate 'Partial Yes', appraisers should check if the authors have performed pilot extraction of data on a sample of the included studies to develop a shared understanding of the data extraction form. 

To rate 'Yes', appraisers should ensure that data is extracted by at least two authors of the review either from all included studies OR from at least a sample of included studies. It is important to check that the review report provides a description of the mechanism used to achieve consensus and shared understanding on which data to collect.

The information of the data extraction process is generally described in the methodology section of the review report.
\newline

\noindent \textbf{Item 7:	Did the authors of the review provide a list of excluded studies, along with the justifications for exclusion, that were read in full text?}
This item refers to studies that were deemed relevant by authors of the review on a reading of the title and abstracts. However, after full-text reading, the authors concluded that the papers are not relevant to the current review. It is expected that the authors of the review should document such papers along with the reason for their exclusion. This will help increase confidence in the results, allow reflecting on the selection criteria used in the study, allow replications, and enable further research (for example, by leveraging on the filtered list of papers for a different analysis).

To rate 'Partial Yes', appraisers should see that the authors of the review have provided a list of such potentially relevant papers. In order to rate 'Yes', justifications for excluding the potentially relevant papers should also be provided. 

This documentation (i.e., a list of potentially relevant papers that were excluded after full-text reading and justifications for excluding them) can be made available in an appendix or as supplementary material for review online (along with other supporting material like the review's protocol).
\newline

\noindent \textbf{Item 8:	Did the authors of the review provide sufficient primary studies’ characteristics to interpret the results?}
The relevance and reliability of a systematic review depends, besides other factors, also on a number of factors related to the included studies such as its type (e.g., case study, survey, and experiment), context (real life or laboratory setting), participants (practitioners or students), and publication venue (e.g., a reputable conference/journal). The review report should describe adequate details about the characteristics of the included studies to inform the review readers about the kind of evidence that is used to draw conclusions.

To rate 'Yes', appraisers should ensure that the authors of the review have provided enough details about the population, interventions (when relevant), outcomes (when relevant), research designs and settings of the included studies.

These details may not be described at one place in a review report, and therefore could be challenging to find. Normally, part of this information is described in the start of the results section in a review report.
\newline

\noindent \textbf{Item 9:	 Did the authors of the review use an appropriate instrument for assessing the quality of primary studies that were included in the review?}
Due to several reasons, including the variety of research designs used in primary studies, reporting quality, use of inconsistent terminology, etc., quality assessment is a challenging task in software engineering systematic literature reviews \cite{kitchenham2015evidence,KITCHENHAM20132049}. Several research-design specific checklists (e.g. for experimentation~\cite{wohlin2012experimentation} and case study research \cite{host2007checklists}) and generic instruments (e.g. \cite{Wieringa12,condori2012experimental}) have been proposed in literature. However, as concluded by Kitchenham \cite{kitchenham2015evidence}, it is not feasible to use the same instrument to assess the quality of different types of studies.

To rate 'Yes', appraiser should ensure that the choice of the instruments used (whether an existing one or one formulated by the authors of the review) has been justified given the goals of the SLR and nature of included studies. Furthermore, the instrument used is expected to evaluate at least the research design, data collection, analysis reporting, and the strength of evidence given the stated goals of the primary study.

The information on the quality assessment of the primary studies is expected to be described in the methodology and results sections of the review report.
\newline

\noindent \textbf{Item 10:	Did the authors of the review use a reliable quality assessment process?}
Like Item 5 and 6, it is important that the quality assessment is not performed solely by a single author of the review.

To rate 'Partial Yes', appraisers should check if at least two authors have performed pilot quality assessment of a sample of the included studies to evaluate the objectivity of the quality assessment criteria and to develop a shared understanding of it. 

To rate 'Yes', appraisers should see that at least two authors of the review independently performed the quality assessment of either all included studies or a sample of included studies (with the remaining performed by one review author) and achieved good consensus. The review report should also describe how differences were resolved in case of different quality scores.

The information about the quality assessment process is typically expected in the methodology section of the review report.
\newline

\noindent \textbf{Item 11:	Were the primary studies appropriately synthesized?}
Synthesis is one of the most important and also challenging parts of a systematic literature review. Without synthesis, the review would be of limited use.

In order to score 'Yes', appraisers need to see if the authors of the review have used and justified an appropriate method for synthesis. It may be the case that the authors of the review do not use the correct or appropriate name for the used synthesis method \cite{CRUZES2011440}. In that case, appraisers would have to carefully read the review report in order to make a decision on this item. The appraisers should further check if the selected synthesis method was appropriately applied and that there is a clear chain of evidence from the answers to the research questions to the data from the primary studies.

The information about the synthesis method and its output may be documented in a separate section. In some cases it may be described in the discussion section after the results section. It could also be the case that the justification for selecting a specific synthesis method is described in the research methodology section, while the outputs of the synthesis step are described in a separate section. 
\newline

\noindent \textbf{Item 12:	Did the authors of the review account for quality of individual studies when interpreting/ discussing the results of the review?}
A review should take the quality of the individual studies into account when interpreting the results. This will increase the confidence in the findings and conclusions of the review.

To rate 'Yes', appraisers should see that either the review has excluded studies that do not meet the quality criteria defined in the study, or the analysis and conclusions are separated based on the quality of the included studies. The information on using the quality of studies while interpreting the results is expected to be described in the discussion or analysis sections of the review report where results are further discussed/analyzed to draw conclusions.
\newline

\noindent \textbf{Item 13:	Did the authors of the review account for primary studies’ characteristics when interpreting/discussing the results of the review?}
There are many factors that can cause heterogeneity in the results of the included studies. It is important to analyze the causes of the heterogeneity in results, if any, while interpreting the results and drawing any conclusions. For example, it could be the variations in the contextual factors (e.g., student versus practitioners as subjects) that lead to differences in the results of different studies. Furthermore, quality scores or some specific quality criteria might also help in explaining the heterogeneity observed in the results \cite{kitchenham2015evidence}. This item is concerned with the use of study characteristics in item 8.

In order to rate 'Yes', appraisers should see that the authors of the review have investigated the impact of study characteristics on the findings of the review.

This discussion is likely to be found after the results section of the review report.
\newline

\noindent \textbf{Item 14:	Did the authors of the review provide appropriate recommendations and conclusions from the review?}
The usefulness of results of the review for the target stakeholders is critical to assess the relevance of the review. This item is a reflection on the aims as motivation for the review assessed in the first item of the instrument (i.e., item 1).

For `Partial Yes' the review should have satisfactory recommendations and conclusions based on the review results. 

For `Yes' the recommendations and conclusions from the review shall be traceable to the review results, clearly targeting specific stakeholders, well aligned with the motivation and provides new insights to the community.

\noindent \textbf{Item 15:	Did the authors of the review report their own potential sources of conflict of interest, including any funding they received for conducting the review?}
To ensure the reliability of a review, it is important that the authors of the review report their sources of funding and any other conflicts of interest. The disclosure of the sources of funding is quite obvious. However, identifying other types of conflicts of interest is not that straightforward. 

For example, if the authors of the review have published on the topic of the review or have a vested interest in the outcome of the review, there is a potential for bias when selecting, analyzing and interpreting their own work and studies with competing alternatives. 

It is encouraged that authors of the review should be experts in the topic area, so it is common that they have published extensively in the topic area. Thus, it is important that the authors of the review document report their effort in identifying any conflicts of interest they have, which is relevant for the review. A mitigation strategy in this case is to establish a protocol and have it reviewed by independent researchers not participating in the literature review.

To rate 'Yes', the appraisers should ensure that the authors of the review have reported on the presence or absence of any conflicts of interest. In case there was some conflict of interest, the authors of the review should have described and justified the steps taken to mitigate the threat of bias in the results of the review.

\end{document}